\documentclass[useAMS,usenatbib]{mn2e}
\usepackage{amsmath}
\usepackage{amsfonts}
\usepackage{xcolor}
\usepackage[draft]{graphicx}

\bibpunct{(}{)}{;}{a}{,}{,}

\makeatletter
\def\ref@jnl#1{{\rmfamily#1}}%
\newcommand\aj{\ref@jnl{AJ}}%
\newcommand\araa{\ref@jnl{ARA\&A}}%
\newcommand\apj{\ref@jnl{ApJ}}%
\newcommand\apjl{\ref@jnl{ApJ}}%
\newcommand\apjs{\ref@jnl{ApJS}}%
\newcommand\ao{\ref@jnl{Appl.~Opt.}}%
\newcommand\apss{\ref@jnl{Ap\&SS}}%
\newcommand\aap{\ref@jnl{A\&A}}%
\newcommand\aapr{\ref@jnl{A\&A~Rev.}}%
\newcommand\aaps{\ref@jnl{A\&AS}}%
\newcommand\azh{\ref@jnl{AZh}}%
\newcommand\baas{\ref@jnl{BAAS}}%
\newcommand\jrasc{\ref@jnl{JRASC}}%
\newcommand\memras{\ref@jnl{MmRAS}}%
\newcommand\mnras{\ref@jnl{MNRAS}}%
\newcommand\pra{\ref@jnl{Phys.~Rev.~A}}%
\newcommand\prb{\ref@jnl{Phys.~Rev.~B}}%
\newcommand\prc{\ref@jnl{Phys.~Rev.~C}}%
\newcommand\prd{\ref@jnl{Phys.~Rev.~D}}%
\newcommand\pre{\ref@jnl{Phys.~Rev.~E}}%
\newcommand\prl{\ref@jnl{Phys.~Rev.~Lett.}}%
\newcommand\pasp{\ref@jnl{PASP}}%
\newcommand\pasj{\ref@jnl{PASJ}}%
\newcommand\qjras{\ref@jnl{QJRAS}}%
\newcommand\skytel{\ref@jnl{S\&T}}%
\newcommand\solphys{\ref@jnl{Sol.~Phys.}}%
\newcommand\sovast{\ref@jnl{Soviet~Ast.}}%
\newcommand\ssr{\ref@jnl{Space~Sci.~Rev.}}%
\newcommand\zap{\ref@jnl{ZAp}}%
\newcommand\nat{\ref@jnl{Nature}}%
\newcommand\iaucirc{\ref@jnl{IAU~Circ.}}%
\newcommand\aplett{\ref@jnl{Astrophys.~Lett.}}%
\newcommand\apspr{\ref@jnl{Astrophys.~Space~Phys.~Res.}}%
\newcommand\bain{\ref@jnl{Bull.~Astron.~Inst.~Netherlands}}%
\newcommand\fcp{\ref@jnl{Fund.~Cosmic~Phys.}}%
\newcommand\gca{\ref@jnl{Geochim.~Cosmochim.~Acta}}%
\newcommand\grl{\ref@jnl{Geophys.~Res.~Lett.}}%
\newcommand\jcp{\ref@jnl{J.~Chem.~Phys.}}%
\newcommand\jgr{\ref@jnl{J.~Geophys.~Res.}}%
\newcommand\jqsrt{\ref@jnl{J.~Quant.~Spec.~Radiat.~Transf.}}%
\newcommand\memsai{\ref@jnl{Mem.~Soc.~Astron.~Italiana}}%
\newcommand\nphysa{\ref@jnl{Nucl.~Phys.~A}}%
\newcommand\physrep{\ref@jnl{Phys.~Rep.}}%
\newcommand\physscr{\ref@jnl{Phys.~Scr}}%
\newcommand\planss{\ref@jnl{Planet.~Space~Sci.}}%
\newcommand\procspie{\ref@jnl{Proc.~SPIE}}%

\long\def\comment#1{}

\def\ba{\begin{eqnarray}}
\def\ea{\end{eqnarray}}
\def\be{\begin{equation}}
\def\ee{\end{equation}}
\def\bdx{\boldsymbol{x }}

\def\bdw{\boldsymbol{w }}

\def\bda{\boldsymbol{a }}
\def\bdn{\boldsymbol{n }}

\def\bdb{\boldsymbol{b }}

\newcommand{\tR}{{\rm R}}
\newcommand{\tA}{{\rm A}}

\def\nobs{N_{\rm obs}}
\def\ncomp{N_{\rm comp}}

\title[CMB and SZ separation with constrained ILC]{CMB and SZ effect separation with Constrained Internal Linear Combinations}
\author[Mathieu Remazeilles, Jacques Delabrouille, Jean-Fran\c{c}ois Cardoso]{Mathieu Remazeilles\thanks{E-mail: remazeil@apc.univ-paris7.fr}, Jacques Delabrouille\thanks{E-mail: delabrouille@apc.univ-paris7.fr}, Jean-Fran\c{c}ois Cardoso\thanks{E-mail: cardoso@enst.fr}\\
APC
10, rue Alice Domon et L\'eonie Duquet,
75205 Paris Cedex 13,
France}

\begin{document}


\pagerange{\pageref{firstpage}--\pageref{lastpage}} \pubyear{2010}

\maketitle

\label{firstpage}

\begin{abstract}
The `Internal Linear Combination' (ILC) component separation method has been extensively used on the data of the WMAP space mission, to extract a single component, the CMB, from the WMAP multifrequency data.
We extend the ILC approach for reconstructing millimeter astrophysical emissions beyond the CMB alone. In particular, we construct a \emph{Constrained} ILC to extract clean maps of both the CMB or the thermal Sunyaev Zeldovich (SZ) effect, with vanishing contamination from the other. The performance of the Constrained ILC is tested on simulations of Planck mission observations, for which we successfully reconstruct independent estimates of the CMB and of the thermal SZ. 
\end{abstract}


\section{Introduction}

The separation of components in observations of the Cosmic Microwave Background (CMB) is an important part of the processing and analysis of such observational data. Various \emph{component separation} methods have been developed to extract the emission of a single component (or of several of them) out of multifrequency observations (see, e.g., \citet{2009LNP...665..159D} for a review). 

Often, such methods assume that the observations are a linear mixture of unknown components (or \emph{sources}), in which case the data are modelled as:
\ba
x_i(p) & = & \sum_{j} A_{ij} s_j(p) + n_i(p)
\label{eq:linear-mixture}
\ea where $x_i(p)$ are $\nobs$ observed maps ($p$ indexing the pixel),
$s_j(p)$ are $\ncomp$ templates for unknown components of interest,
and $n_i(p)$ are maps of noise for the observations. The coefficients
$A_{ij}$ define a \emph{mixing matrix}.

Blind component separation methods such as SMICA
\citep{2003MNRAS.346.1089D,2008ISTSP...2..735C},
FastICA \citep{FastICA,2002MNRAS.334...53M}, JADE \citep{JADE}, CCA
\citep{2006MNRAS.373..271B} or GMCA \citep{2008StMet...5..307B} are primarily designed to solve the problem of estimating the sources $s_i$, separated from one another, in the case where the
observations can be modelled as in equation
\ref{eq:linear-mixture} with the matrix $\tA$ unknown. In practice, the first (and most difficult) task is to estimate $\tA$.

Then, if/when matrix $\tA$ is known (either a priori, or after it has been determined using one of the forementioned blind component separation methods),
the actual component separation is solved by inversion of the linear system of equation
\ref{eq:linear-mixture}. Methods for doing so in the presence of
instrumental noise have been investigated by a number of authors
\citep{1996MNRAS.281.1297T,1999NewA....4..443B,1998MNRAS.300....1H,2002MNRAS.330..807D}.

In CMB observations, in practice, at least one of the columns of the mixing matrix is usually known (i.e. one \emph{mixing vector}) -- that of the CMB. On the other hand, some components cannot be modelled as a single template which is simply scaled by mixing coefficients (e.g. emissions from the galactic interstellar medium). 
Then, the data model of equation \ref{eq:linear-mixture} fails, and one has to resort to other approaches.
For these reasons, the  so--called `Internal Linear Combination' or ILC, wich does not assume any particular parametrisation for foreground emisson, has been extensively used in the analysis of the maps obtained by the WMAP satellite to extract a CMB map \citep{2003ApJS..148...97B,2004ApJ...612..633E,2007ApJ...660..959P,2008arXiv0803.1394K,2009A&A...493..835D}.

The Planck mission, launched on the 14th of May 2009, is a third generation CMB experiment. It observes the sky in 9 frequency channels ranging from 30 to 857 GHz. The high frequency instrument of Planck \citep{2000ApL&C..37..161L,2003NewAR..47.1017L}, in particular, has been designed with bands centred at the minimum, the zero, and the maximum of the thermal SZ emission. The extraction of clean CMB and SZ maps, of a catalog of galaxy clusters selected by their thermal SZ effect, and the investigation of bulk flows in the large scale velocity field from kinetic SZ effect towards galaxy clusters, are part of the scientific programme of Planck.

For these projects, accurate separation of CMB (and kinetic SZ) from
thermal SZ is important
\citep{2001A&A...374....1A,2003JCAP...05..007A}. In particular,
residuals from thermal SZ in the CMB maps can be ``mistaken'' for
detectable kinetic SZ. They can also bias the estimation of
cosmological parameters, in particular those which depend drastically
on the small scale CMB power spectrum, as the scalar spectral index
and the reionisation optical depth \citep{2009MNRAS.392.1153T}.

In the present paper, we address the problem of extending the ILC
method to separate several components of interest with known ``mixing
vector'' in multifrequency observations such as those of Planck. The
method, denoted as \emph{Constrained ILC}, is of interest for
separation of CMB and thermal SZ components with vanising residual
contamination of one by the other.

\section{ILC estimation of CMB and SZ}\label{sec:first-part}

\subsection{Astrophysical components}

Astrophysical components can be separated into two broad categories: diffuse components and point sources. 

Point sources (i.e. unresolved objects) are typically detected and
identified with specific methods, based on spatial filtering (see,
e.g., \citet{2009LNP...665..207B} for a review).  Such methods are
effective when detectable sources are isolated. Experiments observe
also a background of faint sources which, due to limited sensitivity
and resolution, cannot be detected individually.  Such a background is
treated as a diffuse component.
 
Other diffuse components of interest comprise the CMB and diffuse emission from the ISM. SZ effects (thermal  and kinetic), can be considered either as point sources or as diffuse components, depending on the resolution, sensitivity, and frequency coverage of the experiment considered. With Planck, a number of observable clusters will actually be resolved, and we will consider here the thermal SZ as a diffuse component.

The emission of the Interstellar Medium (ISM) is complex, as it involves several processes of emission which are not fully independent. Hot gas emits through synchrotron radiation, warm gas by free-free emission, cold dust by greybody emission, and plausibly by dipole emission from rotating dust grains. Molecules emit through molecular transitions. The dependence of these emissions with frequency is a function of additional parameters which vary over the regions of emission, e.g. electron temperature for free-free, distribution of electrons as a function of energy for the synchrotron, composition and temperature of molecular clouds for molecular line spectra... In addition, as all the ISM is concentrated towards the galactic plane, all of these processes do not result in independent, nor even simply uncorrelated, emissions.

\subsection{The ILC method}

The standard ILC assumes very little about the model of the data. It simply assumes that all available maps ($\nobs$ maps, indexed by $i$) can be written, for all pixels $p$ of the observed maps, as
\begin{equation}
  \label{eq:ILC-model-1comp}
  x_i(p) = a_i s(p) + n_i(p),
\end{equation}
which can be recast as
\begin{equation}
  \label{eq:ILC-model-1comp-vec}
  \bdx(p) =  \bda s(p) +  \bdn(p),
\end{equation}
where $\bdx(p)$ is the vector of observations ($\nobs$ maps), $s(p)$ a
single map of a component of interest, $\bda$ is a known mixing vector
which does not depend on $p$, with as many entries as there are
channels of observation, and $\bdn$ includes instrument noise as well
as all other astrophysical emissions.  For all channels ($\nobs$
maps), it is assumed that all observations are at the same resolution,
although a harmonic or needlet space implementation of the ILC permits
to deal with channel-dependent resolution in a simple way (assuming
symmetric beams).

The ILC provides an estimator $\hat{s}_{\rm ILC}$ of $s$ by forming the linear combination $\hat s(p) = \bdw^t \bdx(p)$ of the observed maps which has unit response to the component of interest (i.e. $\bdw^t\bda = 1$) and has minimum variance. Straightforward algebra leads to ILC coefficients such that $\hat{s}_{\rm ILC}$ of $s$ is given by:
\ba\label{qqq:standardfilter}
\hat{s}_{\rm ILC} & =& \frac{\bda^t \, {\widehat{\tR}}^{-1}}{\bda^t \, {\widehat{\tR}}^{-1} \, \bda} \, \bdx
\label{eq:ILC}
\ea
where ${\widehat{\tR}}$ is the empirical covariance matrix of the observations. 

The ILC component separation method has advantages and drawbacks. The main advantage is that it does not assume a model for the components we are \emph{not} interested in, i.e. all the components whose contributions are collected into a single nuisance term $\bdn(p)$. Drawbacks include the existence of a bias induced by any empirical correlation between the component of interest and the nuisance term, as described in the appendix of \cite{2009A&A...493..835D}, and the need to know the coefficients $a_i$ with some accuracy, especially for sensitive experiments  \citep{2010MNRAS.401.1602D}. 

Note that the ILC relies on the component of interest to be
uncorrelated with the contaminants, i.e. $\langle s(p) n_i(p) \rangle
= 0$ for all channels of observation $i$.

In its simplest implementation, the ILC is performed on the complete
maps, and one single global matrix ${\widehat{\tR}}$ is used. It is
possible, however, to decompose the original maps as sums of different
data subsets, covering each a different region in pixel space or in
harmonic space, to apply independent versions of the ILC to the
different data subsets, and then to recompose a map from all these
independent results. In the present paper, all the experiments done on
simulated maps are performed in harmonic space. We have
${\widehat{\tR}} = {\widehat{\tR}}(\ell)$, and each such covariance
matrix is estimated as
\begin{equation}
  {\widehat{\tR}_{ij}}(\bar{\ell}) 
  = 
  \frac{1}{N_{\bar{\ell}}}\
  \sum_{\bar{\ell}-\Delta\bar{\ell}}^{\bar{\ell}+\Delta\bar{\ell}}~ \sum_m~  x_{\ell m, i}~ x_{\ell m, j}^*
\end{equation}
where $x_{\ell m, i}$ are the spherical harmonic coefficients of map
$x_i(p)$ and where $N_\ell$ is the number of modes $(\ell,m)$ in the
window $[\ell-\Delta\ell,\ell+\Delta\ell]$:
\begin{equation}
  N_{\ell} = \left(\ell+\Delta\ell+1\right)^2 - \left(\ell-\Delta\ell\right)^2.
\end{equation}
Each single ${\widehat{\tR}}(\ell)$ is obtained as the average over a
window in $\ell$.

\subsection{ILC and the SZ effect}

So far, the ILC has been used almost exclusively to extract a CMB map, but it can be used in a similar way to extract any single component which is described as a single template scaling with frequency, provided that the appropriate column of the mixing matrix is known, and that this template is not correlated with other emissions present in the data set.

Sunyaev-Zel'dovich emission \citep{1972CoASP...4..173S}, which can be observed towards clusters of galaxies at millimeter wavelengths, arises through inverse Compton scattering of CMB photons off hot electrons of the intra cluster gas (see, e.g. \citet{1999PhR...310...97B} or \citet{2002ARA&A..40..643C} for a review). 

The thermal SZ emission, in the non-relativistic approximation, is given by the product of a template map $y(p)$ (the map of the cluster Compton parameter, proportional to the integral over the line of sight of $n_e T_e$), and of a known emission law $f(\nu)$. The Compton parameter is given by:
\begin{equation}
  y = \int_{\rm l.o.s.} \frac{kT_e}{m_e c^2} \, n_e \sigma_T \, dl,
\end{equation}
where $k$ is the Boltzmann constant, $m_e$ the electron mass, $c$ the
speed of light, $\sigma_T$ the Thomson cross section, and $n_e$ and
$T_e$ the electron density and temperature respectively. The frequency
dependence of the effect is:
\begin{equation}
\label{eq:SZemlaw}
  f(\nu) =  x(\nu) \frac{e^{x(\nu)}+1}{e^{x(\nu)}-1} - 4
\end{equation}
with
\begin{displaymath}
  x(\nu) = \frac{h\nu}{kT_{\rm CMB}}.
\end{displaymath}
The coefficients $a_i$ of equation \ref{eq:ILC-model-1comp}, for the
thermal SZ, are the integral of the emission law $f(\nu)$ in the
frequency bands of the instrument used to observe the sky.

Kinetic SZ emission has the same emission law as CMB anisotropies, and is proportional to the peculiar velocity of the scattering electron gas (i.e. of the cluster) along the line of sight. Although the kinetic and thermal SZ effects arise from the same set of galaxy clusters, their correlation vanishes because of the sign dependence of the kinetic SZ temperature.

Hence, it is possible to extract a map of the thermal Sunyaev Zel'dovich (SZ) effect from multifrequency observations with an ILC in the same way as is done for the CMB. This method has been implemented and tested in data challenges using simulated data sets, organised in the context of the preparation of the Planck mission \citep{2008A&A...491..597L,Challenge-SZ}. 

Note that on the other hand, a CMB map obtained by an ILC contains
emission from \emph{both} primary CMB anisotropies and kinetic SZ
effect. The latter is typically very small compared to primary CMB
anisotropies. For high resolution high sensitivity maps however, the
kinetic SZ can be separated from the CMB by matched filtering, using
the cluster profile estimated from the thermal SZ map.

\subsection{A two-component model for the ILC}

Without assuming much about the detailed properties of the emissions
of other foregrounds, and supposing we are interested mostly in the
CMB and the SZ, both thermal (tSZ) and kinetic (kSZ), the
observational data can be modeled by an extension of equations
\ref{eq:ILC-model-1comp} and \ref{eq:ILC-model-1comp-vec}, as: 
\begin{equation}
  \label{eq:ILC-model-2comp}
  x_i(p)= a_i s(p)+b_i y(p)+ n_i(p)  
\end{equation}
or as
\begin{equation}
  \bdx (p)= \bda s(p) + \bdb y(p) + \bdn (p)  
\end{equation}
where, as in equation~\ref{eq:ILC-model-1comp}, $s(p)$ is the CMB map
(including kSZ), but where now $y(p)$ is the thermal SZ map and $\bdn =
\{n_i(p)\}$ includes both instrumental noise and unmodeled
astrophysical foregrounds (i.e. all sky components except CMB, kSZ and tSZ), in all frequency channels.  Vector $\bda =
(1,1,...,1)^t$ is the CMB mixing vector and $\bdb$ is the mixing
vector of the thermal SZ, as derived from Eq.~(\ref{eq:SZemlaw}).

\subsection{Independent ILC estimation of CMB and SZ}

It is a straightforward process to reconstruct both a (CMB+kSZ) and a
tSZ map independently by implementing two separate ILC, one for the
(CMB+kSZ), and one for the tSZ.  Standard ILC gives:
\begin{displaymath}
  \hat{s}(p) = \frac{\bda^t \, {\widehat{\tR}}^{-1}}{\bda^t \, {\widehat{\tR}}^{-1} \, \bda} \, \bdx(p),
  \qquad
  \hat{y}(p) = \frac{\bdb^t \, {\widehat{\tR}}^{-1}}{\bdb^t \, {\widehat{\tR}}^{-1} \, \bdb} \, \bdx(p).  
\end{displaymath}
Those estimates, however, do not take fully into account the prior
knowledge about the existence of two components entering in the
observations as described by equation~\ref{eq:ILC-model-2comp}.
In particular, the ILC weights used for CMB reconstruction do not
guarantee that the reconstructed CMB contains no thermal
SZ. Similarly, the weights used to reconstruct the thermal SZ do not
guarantee that the reconstructed thermal SZ does not contain any CMB
(and kinetic SZ). In fact, using the solutions above with the data
model of equation \ref{eq:ILC-model-2comp}, we have: 
\begin{displaymath}
  \hat{s}(p) =
  s(p) + \frac{\bda^t \, {\widehat{\tR}}^{-1} \bdb}{\bda^t \,
    {\widehat{\tR}}^{-1} \, \bda} \, y(p) + \frac{\bda^t \,
    {\widehat{\tR}}^{-1}}{\bda^t \, {\widehat{\tR}}^{-1} \, \bda} \,
  \bdn(p) 
\end{displaymath}
The second term on the right hand side is the contamination of the
recovered (CMB+kSZ) by thermal SZ. Similarly, when one recovers a tSZ
map, we get:
\begin{displaymath}
  \hat{y}(p) = y(p) + \frac{\bdb^t \,
    {\widehat{\tR}}^{-1} \bda}{\bdb^t \, {\widehat{\tR}}^{-1} \, \bdb}
  \, s(p) + \frac{\bdb^t \, {\widehat{\tR}}^{-1}}{\bdb^t \,
    {\widehat{\tR}}^{-1} \, \bdb} \, \bdn(p) .
\end{displaymath}
These solutions minimise (by construction) the total variance of the
reconstructed maps, but both maps contain contamination from the other
component of interest. For certain applications, this contamination is
not acceptable. For instance, significant thermal SZ leaking into the
reconstructed CMB will make it difficult to extract any kinetic SZ
information from the reconstructed (CMB+kSZ) map.

\begin{figure*}
  \begin{center}
    \includegraphics[width=5.3cm]{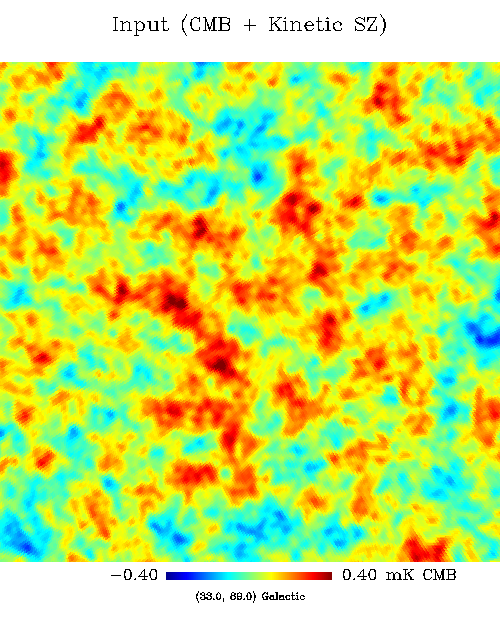}
    \includegraphics[width=5.3cm]{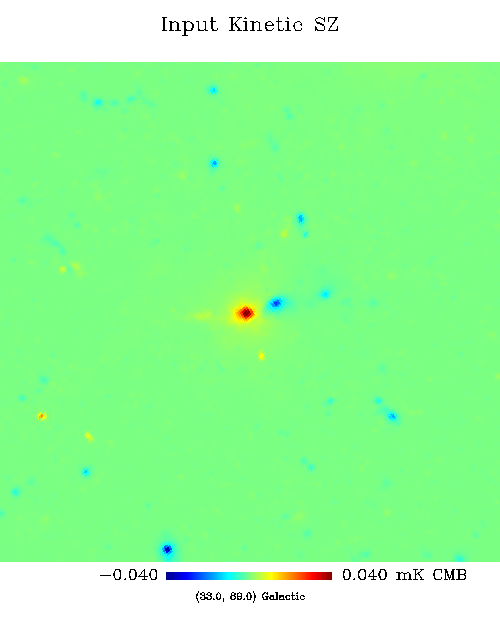}
    \includegraphics[width=5.3cm]{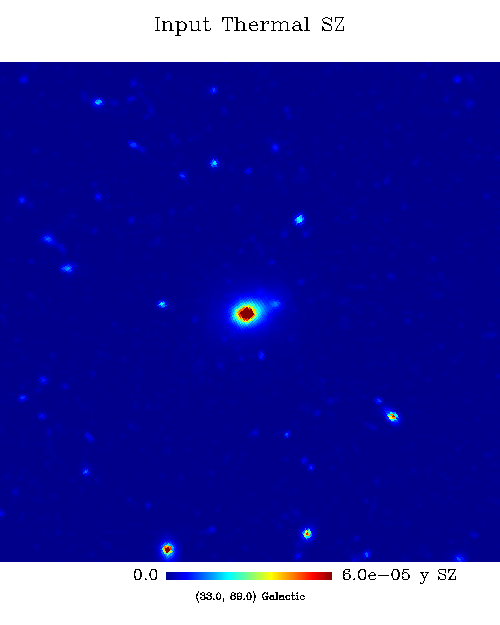}
    \includegraphics[width=5.3cm]{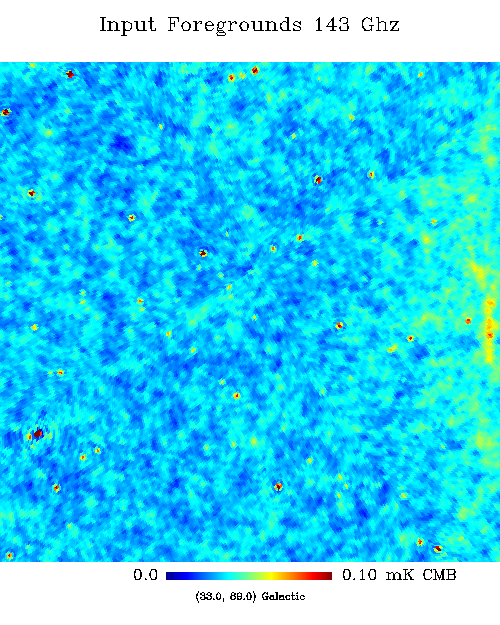}
    \includegraphics[width=5.3cm]{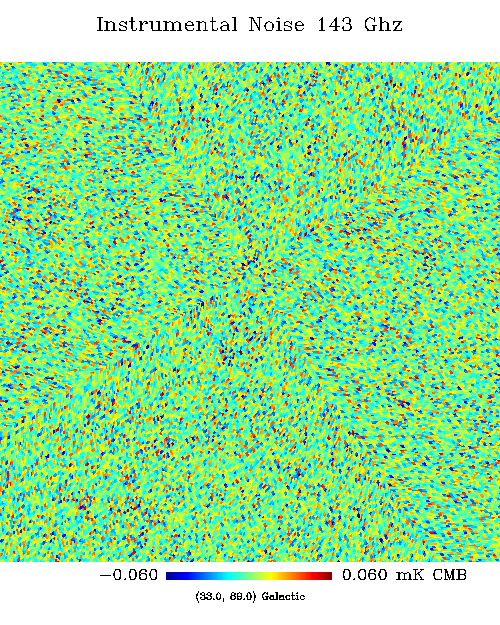}
    \includegraphics[width=5.3cm]{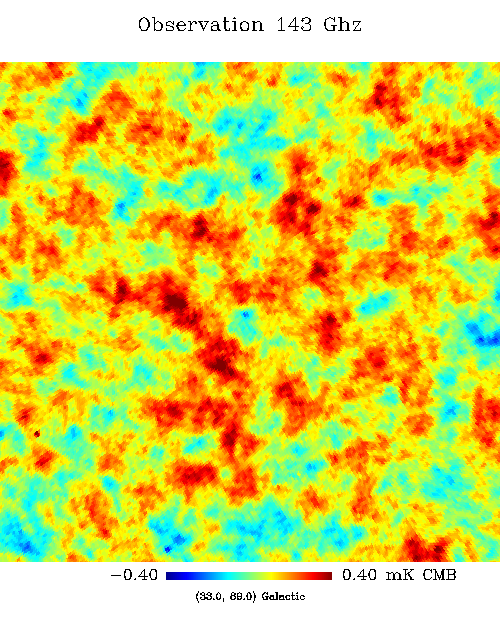}
  \end{center}
\caption{{\bf Simulated observations:} A patch of the simulated sky located at high galactic latitude, around galactic coordinates of ${(l,b) = (33^\circ,89^\circ)}$. From top left to bottom right: CMB (including kinetic SZ), Kinetic SZ, Thermal SZ, other foregrounds (at 143 GHz), Noise of the 143 GHz channel, Total observed emission (including noise) at 143 GHz.}
\label{Fig:inputs}
\end{figure*}

\begin{figure*}
  \begin{center}
    \includegraphics[width=5.3cm]{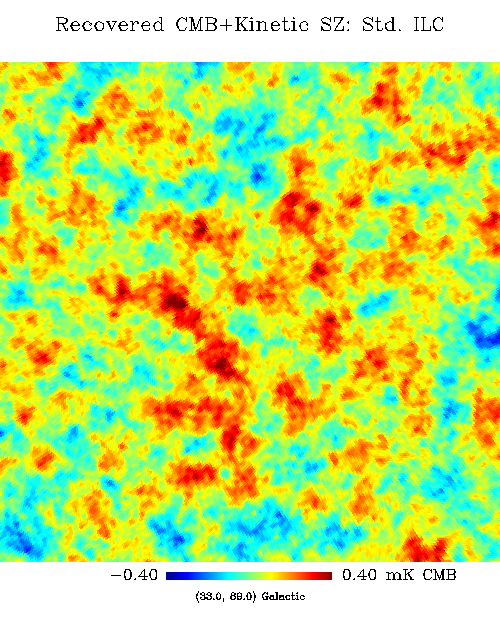}
    \includegraphics[width=5.3cm]{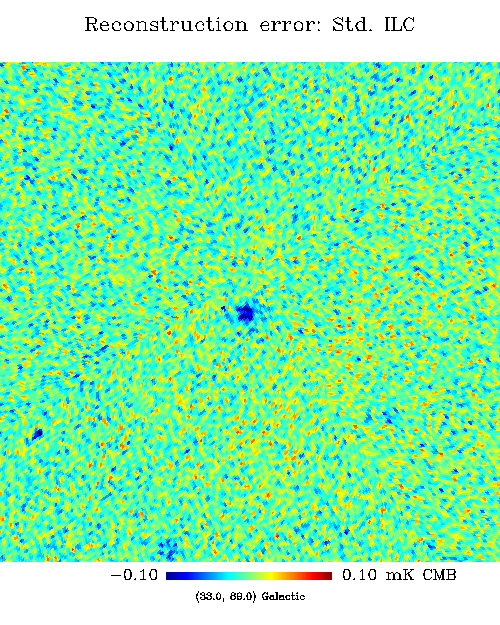}
    \includegraphics[width=5.3cm]{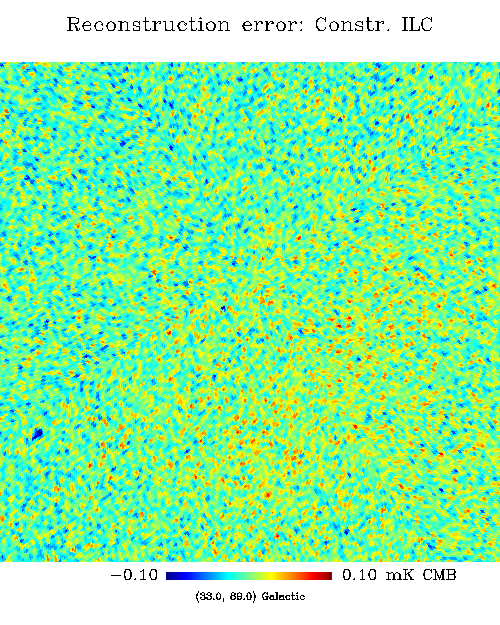}
  \end{center}
\caption{{\bf CMB extraction}: Left panel: The CMB reconstructed by standard ILC from simulated Planck observations (visualisation of the small patch of figure \ref{Fig:inputs}). Reconstruction looks excellent. Middle panel: a detailed examination of the reconstruction error (recovered CMB minus input CMB, including the kSZ), reveals dark patches towards the direction of galaxy clusters. This is due to the leakage of thermal SZ into the reconstructed CMB. Right panel: a constrained ILC guarantees the absence of contaminating tSZ in the reconstructed map, with minimal impact on the total error variance.
}
\label{Fig:cmb}
\end{figure*}

\subsection{Constrained ILC estimation of CMB and SZ}\label{subsec:constrained-filter}

Now assume that we want to ensure that each reconstructed map of interest contains no contamination from the other component, i.e. the CMB map contains no contribution from the thermal SZ, and the thermal SZ no contribution from the CMB (nor from kinetic SZ). 

In the spirit of the ILC, we then look for a minimum variance estimate
$\hat{s}$ of the CMB map\footnote{The same derivation can be performed
  for thermal SZ reconstruction, by simple inversion of the roles of
  $\bda$ and $\bdb$ and of $s$ and~$y$.} $s$ as a linear combination
of the frequency maps $x_i$:
\begin{displaymath}
  \hat{s} = \bdw^t \bdx
\end{displaymath}
where the weights $w_i$ have to satisfy the constraints:
\begin{align}
  \bdw^t \bda &=  1,\label{qqq:c1}\\
  \bdw^t \bdb &=  0,\label{qqq:c2} 
\end{align}
so that we conserve the CMB signal (including kSZ) and completely
eliminate the thermal SZ signal.  
The weights thus are solutions of
\begin{displaymath}
  \frac{\partial}{\partial
    w_i}\left[\bdw^t \widehat{\tR} \bdw + \lambda\left(1-\bdw^t
      \bda\right)-\mu \bdw^t \bdb\right] = 0 
\end{displaymath}
where $\widehat{\tR}$ is the empirical covariance matrix of the
observed maps and $\lambda, \mu$ are Lagrange multipliers. 
We find the solution: 
\begin{equation}
  \label{qqq:sol}
  \bdw = \lambda  \widehat{\tR}^{-1}\bda + \mu \widehat{\tR}^{-1}\bdb.    
\end{equation}
Applying constraints (\ref{qqq:c1}), (\ref{qqq:c2}) to the solution
(\ref{qqq:sol}), we have to solve the system 
\begin{align}
  \lambda \bda^t\widehat{\tR}^{-1}\bda + \mu \bda^t\widehat{\tR}^{-1}\bdb &=1,\\
  \lambda \bdb^t\widehat{\tR}^{-1}\bda + \mu \bdb^t\widehat{\tR}^{-1}\bdb &=0.
\end{align}
If $\bdb$ is not proportional to $\bda$ (which is guaranteed since the
CMB and the thermal SZ do not have the same emission law), there is a
unique solution to the system, for which the Lagrange multipliers are:
\begin{align}
  \lambda &= {\bdb^t\widehat{\tR}^{-1}\bdb
    \over \left(\bda^t\widehat{\tR}^{-1}\bda\right)\left(\bdb^t\widehat{\tR}^{-1}\bdb\right)-\left(\bda^t\widehat{\tR}^{-1}\bdb\right)^2},\\
  \mu &= - {\bda^t\widehat{\tR}^{-1}\bdb\over
    \left(\bda^t\widehat{\tR}^{-1}\bda\right)\left(\bdb^t\widehat{\tR}^{-1}\bdb\right)-\left(\bda^t\widehat{\tR}^{-1}\bdb\right)^2}.
\end{align}
The weights of the constrained ILC (for CMB+kSZ reconstruction) are
given by 
\begin{equation}
  \label{eq:ILC-2comp}
  \bdw^t =
  {\left(\bdb^t\widehat{\tR}^{-1}\bdb\right)\bda^t \widehat{\tR}^{-1} -
    \left(\bda^t\widehat{\tR}^{-1}\bdb\right)\bdb^t
    \widehat{\tR}^{-1}\over
    \left(\bda^t\widehat{\tR}^{-1}\bda\right)\left(\bdb^t\widehat{\tR}^{-1}\bdb\right)-\left(\bda^t\widehat{\tR}^{-1}\bdb\right)^2}.
\end{equation}
A similar expression with the role of $\bda$ and $\bdb$ exchanged
holds for the weights to be used for estimating the thermal SZ map
$y(p)$.

This is the generalisation of the ILC component separation when two
components are recovered, and when one imposes that there is no
leakage of one component in the reconstructed map of the other.

\subsection{Application to simulated Planck observations}\label{subsec:psm}

We investigate the performance of standard ILC and constrained ILC for separating the CMB and the thermal SZ using observations such as those of the Planck mission.

\subsubsection{Simulations}

Our investigations are carried out on sky simulations generated with the Planck Sky Model (PSM) version 1.6.3 for all Planck LFI and HFI channels. Sky simulations include Gaussian CMB generated assuming a $C_\ell$ model fitting the WMAP 5 year observations \citep{2009ApJS..180..225H}, thermal and kinetic SZ effect, four components of galactic ISM emission including thermal and spinning dust, synchrotron, and free-free, and emission from point sources (radio and infrared). The resolution and noise level of the observations correspond to nominal mission parameters as described in the Planck `Blue Book'. Some details about PSM simulations can be found in \cite{2008A&A...491..597L} and \cite{2009A&A...503..691B}.
Figure \ref{Fig:inputs} displays the simulated emission in a small patch centred around galactic coordinates of $(l,b) = (33^\circ,89^\circ)$ (selected, in our full sky maps, for the presence of a bright galaxy cluster at high galactic latitude, where galactic foregrounds are low, which permits to illustrate best the separation of CMB and SZ).

\subsubsection{ILC results: the CMB}

As a first step, we implement the usual single component ILC (standard
ILC) for both the CMB and the thermal SZ. For this particular
application, a small galactic mask is used to blank out the region of
strongest galactic emission, which permits to implement the ILC in
harmonic space (i.e priority is given to the localisation of the
filters in harmonic space, rather than pixel space).  The spectral
statistics ${\widehat{\tR}}(\ell)$ are computed in windows of $\ell$
of width $\Delta \ell = 50$ at low $\ell$, and $\Delta \ell = 20$ for
the highest multipoles. The ILC is performed independently for each
$\ell$.

Figure \ref{Fig:cmb} displays the results of the reconstruction of the
CMB.  The left panel shows the standard ILC reconstruction of the CMB
map, which can hardly be distinguished from the left panel of figure
\ref{Fig:inputs} (the reconstruction is visually excellent, albeit one
may notice faint small scale granularity caused by noise). The middle
panel shows the error (difference input-output) map using a standard
harmonic domain ILC. Negative patches, which could not be picked out
in the reconstructed CMB map, are clearly seen in the direction of the
brightest clusters : the standard ILC, clearly, does not reject
perfectly the thermal SZ effect.  The amplitude of the thermal SZ
leaking into the map is of 0.1 mK (thermodynamic) for the brightest
cluster in the field, about 2.5 times the kinetic SZ effect for that
particular galaxy cluster (and well above the Planck noise level).

The right panel shows the reconstruction error when the constrained
ILC is used instead (the corresponding CMB map cannot be distinguished
by visual inspection from what is obtained with the standard ILC, and
is not displayed.  As expected, there is no indication in the error
map of leakage of thermal SZ into the reconstructed map of CMB +
kinetic SZ.  In the present case, the impact of the additional
constraint of vanishing thermal SZ contribution in the CMB has
negligible impact on the total noise level.

That example demonstrates that with data sets such as those of Planck,
constraining the ILC to reject the thermal SZ effect makes it possible
to avoid contaminating the CMB with SZ, with very little impact on the
overall level of noise in the reconstructed CMB.  This statement,
however, does not hold for any possible experiment.  In the case of
observations with more noise and less frequency channels, it may well
be that constraining the ILC would actually result in significantly
increased noise level, for negligible gain on the contamination by SZ.

\subsubsection{ILC results: the SZ effect}

\begin{figure}
  \begin{center}
    \includegraphics[width=5.3cm]{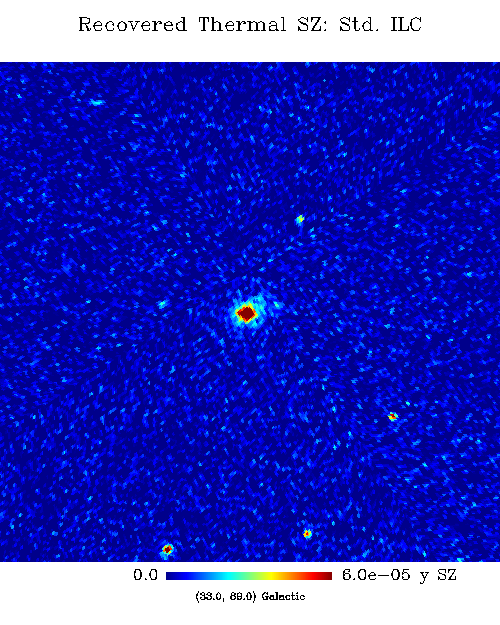}
  \end{center}
\caption{{Thermal SZ standard ILC reconstruction centred around galactic coordinates of ${(l,b) = (33^\circ,89^\circ)}$ (${5'}$ resolution).}
 }
\label{Fig:sz}
\end{figure}

Similarly, we show in figure~\ref{Fig:sz} the reconstruction of
thermal SZ effect by a standard ILC in that same area of the sky.
Clusters are clearly visible, which confirms the sensitivity of Planck
for detecting galaxy clusters and doing SZ science. It is clear that
the very specific SZ emission law, with negative signal below 217 GHz
and positive signal at higher frequencies, helps to separate it from
other emissions effectively.

We compare the performance of the Standard ILC result (shown in figure
\ref{Fig:sz}), and a constrained ILC reconstruction (not shown), where
the solution is constrained to perfectly reject components with the
same color as that of CMB anisotropies (\textit{i.e.} both the CMB and
the kinetic SZ).  The result of the constrained ILC is not displayed
since it is visually indistinguishable from the standard ILC result.
Visual inspection of reconstruction errors (reconstructed thermal SZ
minus true input thermal SZ at the same resolution) does not reveal
any particular feature connected to any astrophysical component, in
either reconstructed SZ map.

Hence, in the case of the reconstruction of the thermal SZ, the
constrained ILC result is very similar to the standard one. The main
reason for this lack of qualitative difference between the two filters
is simple: as the CMB is seen by all channels with very good signal to
noise ratio, the standard ILC always adjusts the ILC weights to
null-out the CMB (or nearly so). Little is gained by imposing this
constraint explicitely \emph{a priori}.

\section{Conclusion}

In this article we have developed a \emph{Constrained ILC} method, and have shown how it can be used to reconstruct the CMB and the thermal SZ component with vanishing contamination of one by the other.
We have applied the filters to simulations of Planck observations, and have shown that they permit to reconstruct both a CMB and a thermal SZ map with excellent performance.

\section*{Acknowledgements}
\thanks{
Some of the results in this paper have been derived using the
HEALPix package \citep{gorski05}.  We acknowledge the use of the Planck
Sky Model, developed by the Component Separation Working Group (WG2) of the
Planck Collaboration.}

\bibliography{constrained_ilc}

\begin{thebibliography}{32}
\expandafter\ifx\csname natexlab\endcsname\relax\def\natexlab#1{#1}\fi

\bibitem[{{Aghanim} {et~al.}(2001){Aghanim}, {G{\'o}rski}, \&
  {Puget}}]{2001A&A...374....1A}
{Aghanim}, N., {G{\'o}rski}, K.~M., \& {Puget}, J. 2001, \aap, 374, 1

\bibitem[{{Aghanim} {et~al.}(2003){Aghanim}, {Hansen}, {Pastor}, \&
  {Semikoz}}]{2003JCAP...05..007A}
{Aghanim}, N., {Hansen}, S.~H., {Pastor}, S., \& {Semikoz}, D.~V. 2003, Journal
  of Cosmology and Astro-Particle Physics, 5, 7

\bibitem[{{Barreiro}(2009)}]{2009LNP...665..207B}
{Barreiro}, R.~B. 2009, in Lecture Notes in Physics, Berlin Springer Verlag,
  Vol. 665, Data Analysis in Cosmology, ed. {V.~J.~Martinez, E.~Saar,
  E.~M.~Gonzales, \& M.~J.~Pons-Borderia }, 207--+

\bibitem[{{Bennett} {et~al.}(2003){Bennett}, {Hill}, {Hinshaw}, {Nolta},
  {Odegard}, {Page}, {Spergel}, {Weiland}, {Wright}, {Halpern}, {Jarosik},
  {Kogut}, {Limon}, {Meyer}, {Tucker}, \& {Wollack}}]{2003ApJS..148...97B}
{Bennett}, C.~L., {Hill}, R.~S., {Hinshaw}, G., {Nolta}, M.~R., {Odegard}, N.,
  {Page}, L., {Spergel}, D.~N., {Weiland}, J.~L., {Wright}, E.~L., {Halpern},
  M., {Jarosik}, N., {Kogut}, A., {Limon}, M., {Meyer}, S.~S., {Tucker}, G.~S.,
  \& {Wollack}, E. 2003, \apjs, 148, 97

\bibitem[{{Betoule} {et~al.}(2009){Betoule}, {Pierpaoli}, {Delabrouille}, {Le
  Jeune}, \& {Cardoso}}]{2009A&A...503..691B}
{Betoule}, M., {Pierpaoli}, E., {Delabrouille}, J., {Le Jeune}, M., \&
  {Cardoso}, J. 2009, \aap, 503, 691

\bibitem[{{Birkinshaw}(1999)}]{1999PhR...310...97B}
{Birkinshaw}, M. 1999, \physrep, 310, 97

\bibitem[{{Bobin} {et~al.}(2008){Bobin}, {Moudden}, {Starck}, {Fadili}, \&
  {Aghanim}}]{2008StMet...5..307B}
{Bobin}, J., {Moudden}, Y., {Starck}, J., {Fadili}, J., \& {Aghanim}, N. 2008,
  Statistical Methodology, 5, 307

\bibitem[{{Bonaldi} {et~al.}(2006){Bonaldi}, {Bedini}, {Salerno},
  {Baccigalupi}, \& {de Zotti}}]{2006MNRAS.373..271B}
{Bonaldi}, A., {Bedini}, L., {Salerno}, E., {Baccigalupi}, C., \& {de Zotti},
  G. 2006, \mnras, 373, 271

\bibitem[{{Bouchet} \& {Gispert}(1999)}]{1999NewA....4..443B}
{Bouchet}, F.~R., \& {Gispert}, R. 1999, New Astronomy, 4, 443

\bibitem[{{Cardoso}(1998)}]{JADE}
{Cardoso}, J.-F. 1998, Proceedings of the IEEE, 9, 2009

\bibitem[{{Cardoso} {et~al.}(2008){Cardoso}, {Le Jeune}, {Delabrouille},
  {Betoule}, \& {Patanchon}}]{2008ISTSP...2..735C}
{Cardoso}, J.-F., {Le Jeune}, M., {Delabrouille}, J., {Betoule}, M., \&
  {Patanchon}, G. 2008, IEEE Journal of Selected Topics in Signal Processing,
  vol.~2, issue 5, pp.~735-746, 2, 735

\bibitem[{{Carlstrom} {et~al.}(2002){Carlstrom}, {Holder}, \&
  {Reese}}]{2002ARA&A..40..643C}
{Carlstrom}, J.~E., {Holder}, G.~P., \& {Reese}, E.~D. 2002, \araa, 40, 643

\bibitem[{{Delabrouille} \& {Cardoso}(2009)}]{2009LNP...665..159D}
{Delabrouille}, J., \& {Cardoso}, J. 2009, in Lecture Notes in Physics, Berlin
  Springer Verlag, Vol. 665, Lecture Notes in Physics, Berlin Springer Verlag,
  ed. {V.~J.~Martinez, E.~Saar, E.~M.~Gonzales, \& M.~J.~Pons-Borderia },
  159--+

\bibitem[{{Delabrouille} {et~al.}(2009){Delabrouille}, {Cardoso}, {Le Jeune},
  {Betoule}, {Fay}, \& {Guilloux}}]{2009A&A...493..835D}
{Delabrouille}, J., {Cardoso}, J., {Le Jeune}, M., {Betoule}, M., {Fay}, G., \&
  {Guilloux}, F. 2009, \aap, 493, 835

\bibitem[{{Delabrouille} {et~al.}(2003){Delabrouille}, {Cardoso}, \&
  {Patanchon}}]{2003MNRAS.346.1089D}
{Delabrouille}, J., {Cardoso}, J., \& {Patanchon}, G. 2003, \mnras, 346, 1089

\bibitem[{{Delabrouille} {et~al.}(2002){Delabrouille}, {Patanchon}, \&
  {Audit}}]{2002MNRAS.330..807D}
{Delabrouille}, J., {Patanchon}, G., \& {Audit}, E. 2002, \mnras, 330, 807

\bibitem[{{Dick} {et~al.}(2010){Dick}, {Remazeilles}, \&
  {Delabrouille}}]{2010MNRAS.401.1602D}
{Dick}, J., {Remazeilles}, M., \& {Delabrouille}, J. 2010, \mnras, 401, 1602

\bibitem[{{Eriksen} {et~al.}(2004){Eriksen}, {Banday}, {G{\'o}rski}, \&
  {Lilje}}]{2004ApJ...612..633E}
{Eriksen}, H.~K., {Banday}, A.~J., {G{\'o}rski}, K.~M., \& {Lilje}, P.~B. 2004,
  \apj, 612, 633

\bibitem[{{G{\'o}rski} {et~al.}(2005){G{\'o}rski}, {Hivon}, {Banday},
  {Wandelt}, {Hansen}, {Reinecke}, \& {Bartelmann}}]{gorski05}
{G{\'o}rski}, K.~M., {Hivon}, E., {Banday}, A.~J., {Wandelt}, B.~D., {Hansen},
  F.~K., {Reinecke}, M., \& {Bartelmann}, M. 2005, \apj, 622, 759

\bibitem[{{Hinshaw} {et~al.}(2009){Hinshaw}, {Weiland}, {Hill}, {Odegard},
  {Larson}, {Bennett}, {Dunkley}, {Gold}, {Greason}, {Jarosik}, {Komatsu},
  {Nolta}, {Page}, {Spergel}, {Wollack}, {Halpern}, {Kogut}, {Limon}, {Meyer},
  {Tucker}, \& {Wright}}]{2009ApJS..180..225H}
{Hinshaw}, G., {Weiland}, J.~L., {Hill}, R.~S., {Odegard}, N., {Larson}, D.,
  {Bennett}, C.~L., {Dunkley}, J., {Gold}, B., {Greason}, M.~R., {Jarosik}, N.,
  {Komatsu}, E., {Nolta}, M.~R., {Page}, L., {Spergel}, D.~N., {Wollack}, E.,
  {Halpern}, M., {Kogut}, A., {Limon}, M., {Meyer}, S.~S., {Tucker}, G.~S., \&
  {Wright}, E.~L. 2009, \apjs, 180, 225

\bibitem[{{Hobson} {et~al.}(1998){Hobson}, {Jones}, {Lasenby}, \&
  {Bouchet}}]{1998MNRAS.300....1H}
{Hobson}, M.~P., {Jones}, A.~W., {Lasenby}, A.~N., \& {Bouchet}, F.~R. 1998,
  \mnras, 300, 1

\bibitem[{{Hyvarinen}(1999)}]{FastICA}
{Hyvarinen}, A. 1999, IEEE Transactions on Neural Networks, 10, 626

\bibitem[{{Kim} {et~al.}(2008){Kim}, {Naselsky}, \&
  {Christensen}}]{2008arXiv0803.1394K}
{Kim}, J., {Naselsky}, P., \& {Christensen}, P.~R. 2008, ArXiv e-prints, 803

\bibitem[{{Lamarre} {et~al.}(2000){Lamarre}, {Ade}, {Beno{\^i}t}, {de
  Bernardis}, {Bock}, {Bouchet}, {Bradshaw}, {Charra}, {Church}, {Couchot},
  {Delabrouille}, {Efstathiou}, {Giard}, {Giraud-H{\'e}raud}, {Gispert},
  {Griffin}, {Lange}, {Murphy}, {Pajot}, {Puget}, \&
  {Ristorcelli}}]{2000ApL&C..37..161L}
{Lamarre}, J.-M., {Ade}, P.~R., {Beno{\^i}t}, A., {de Bernardis}, P., {Bock},
  J., {Bouchet}, F., {Bradshaw}, T., {Charra}, J., {Church}, S., {Couchot}, F.,
  {Delabrouille}, J., {Efstathiou}, G., {Giard}, M., {Giraud-H{\'e}raud}, Y.,
  {Gispert}, R., {Griffin}, M., {Lange}, A., {Murphy}, A., {Pajot}, F.,
  {Puget}, J.~L., \& {Ristorcelli}, I. 2000, Astrophysical Letters
  Communications, 37, 161

\bibitem[{{Lamarre} {et~al.}(2003){Lamarre}, {Puget}, {Bouchet}, {Ade},
  {Benoit}, {Bernard}, {Bock}, {de Bernardis}, {Charra}, {Couchot},
  {Delabrouille}, {Efstathiou}, {Giard}, {Guyot}, {Lange}, {Maffei}, {Murphy},
  {Pajot}, {Piat}, {Ristorcelli}, {Santos}, {Sudiwala}, {Sygnet}, {Torre},
  {Yurchenko}, \& {Yvon}}]{2003NewAR..47.1017L}
{Lamarre}, J.-M., {Puget}, J.~L., {Bouchet}, F., {Ade}, P.~A.~R., {Benoit}, A.,
  {Bernard}, J.~P., {Bock}, J., {de Bernardis}, P., {Charra}, J., {Couchot},
  F., {Delabrouille}, J., {Efstathiou}, G., {Giard}, M., {Guyot}, G., {Lange},
  A., {Maffei}, B., {Murphy}, A., {Pajot}, F., {Piat}, M., {Ristorcelli}, I.,
  {Santos}, D., {Sudiwala}, R., {Sygnet}, J.~F., {Torre}, J.~P., {Yurchenko},
  V., \& {Yvon}, D. 2003, New Astronomy Review, 47, 1017

\bibitem[{{Leach} {et~al.}(2008){Leach}, {Cardoso}, {Baccigalupi}, {Barreiro},
  {Betoule}, {Bobin}, {Bonaldi}, {Delabrouille}, {de Zotti}, {Dickinson},
  {Eriksen}, {Gonz{\'a}lez-Nuevo}, {Hansen}, {Herranz}, {Le Jeune},
  {L{\'o}pez-Caniego}, {Mart{\'{\i}}nez-Gonz{\'a}lez}, {Massardi}, {Melin},
  {Miville-Desch{\^e}nes}, {Patanchon}, {Prunet}, {Ricciardi}, {Salerno},
  {Sanz}, {Starck}, {Stivoli}, {Stolyarov}, {Stompor}, \&
  {Vielva}}]{2008A&A...491..597L}
{Leach}, S.~M., {Cardoso}, J.-F., {Baccigalupi}, C., {Barreiro}, R.~B.,
  {Betoule}, M., {Bobin}, J., {Bonaldi}, A., {Delabrouille}, J., {de Zotti},
  G., {Dickinson}, C., {Eriksen}, H.~K., {Gonz{\'a}lez-Nuevo}, J., {Hansen},
  F.~K., {Herranz}, D., {Le Jeune}, M., {L{\'o}pez-Caniego}, M.,
  {Mart{\'{\i}}nez-Gonz{\'a}lez}, E., {Massardi}, M., {Melin}, J.,
  {Miville-Desch{\^e}nes}, M., {Patanchon}, G., {Prunet}, S., {Ricciardi}, S.,
  {Salerno}, E., {Sanz}, J.~L., {Starck}, J., {Stivoli}, F., {Stolyarov}, V.,
  {Stompor}, R., \& {Vielva}, P. 2008, \aap, 491, 597

\bibitem[{{Maino} {et~al.}(2002){Maino}, {Farusi}, {Baccigalupi}, {Perrotta},
  {Banday}, {Bedini}, {Burigana}, {De Zotti}, {G{\'o}rski}, \&
  {Salerno}}]{2002MNRAS.334...53M}
{Maino}, D., {Farusi}, A., {Baccigalupi}, C., {Perrotta}, F., {Banday}, A.~J.,
  {Bedini}, L., {Burigana}, C., {De Zotti}, G., {G{\'o}rski}, K.~M., \&
  {Salerno}, E. 2002, \mnras, 334, 53

\bibitem[{{Melin} {et~al.}(2010){Melin}, {Aghanim}, {Bartelmann}, {Bartlett},
  {Betoule}, Bobin, Carvalho, Chon, Delabrouille, Diego, Harrison, Herranz,
  Hobson, Kneissl, Lasenby, Le~Jeune, Lopez-Caniego, Mazzotta, Schaefer,
  Starck, Waizmann, \& Yvon}]{Challenge-SZ}
{Melin}, J.-B., {Aghanim}, N., {Bartelmann}, M., {Bartlett}, J.~G., {Betoule},
  M., Bobin, J., Carvalho, P., Chon, G., Delabrouille, J., Diego, J.~M.,
  Harrison, D.~L., Herranz, D., Hobson, M., Kneissl, R., Lasenby, A.~N.,
  Le~Jeune, M., Lopez-Caniego, M., Mazzotta, P., Schaefer, B., Starck, J.-L.,
  Waizmann, J.-C., \& Yvon, D. 2010, in preparation

\bibitem[{{Park} {et~al.}(2007){Park}, {Park}, \& {Gott}}]{2007ApJ...660..959P}
{Park}, C.-G., {Park}, C., \& {Gott}, J.~R.~I. 2007, \apj, 660, 959

\bibitem[{{Sunyaev} \& {Zeldovich}(1972)}]{1972CoASP...4..173S}
{Sunyaev}, R.~A., \& {Zeldovich}, Y.~B. 1972, Comments on Astrophysics and
  Space Physics, 4, 173

\bibitem[{{Taburet} {et~al.}(2009){Taburet}, {Aghanim}, {Douspis}, \&
  {Langer}}]{2009MNRAS.392.1153T}
{Taburet}, N., {Aghanim}, N., {Douspis}, M., \& {Langer}, M. 2009, \mnras, 392,
  1153

\bibitem[{{Tegmark} \& {Efstathiou}(1996)}]{1996MNRAS.281.1297T}
{Tegmark}, M., \& {Efstathiou}, G. 1996, \mnras, 281, 1297

\end{thebibliography}

\label{lastpage}

\end{document}